\documentclass[conference]{IEEEtran}
\IEEEoverridecommandlockouts

\usepackage[russian, ngerman, english]{babel}
\usepackage[utf8]{inputenc}
\usepackage[T2A,T1]{fontenc}

\usepackage{anyfontsize}
\usepackage{csquotes}
\usepackage{amsmath,amssymb,amsfonts}
\usepackage{algorithmic}
\usepackage{graphicx}
\usepackage{textcomp}
\usepackage{xcolor}
\usepackage{orcidlink}
\usepackage{array}

\usepackage{tikz}
\usetikzlibrary{positioning}
\usepackage{pgfplots}
\DeclareUnicodeCharacter{2212}{−}
\usepgfplotslibrary{groupplots,dateplot}
\usetikzlibrary{patterns,shapes.arrows}
\pgfplotsset{compat=newest}

\usepackage[style=ieee]{biblatex} 

\DeclareMathOperator*{\argmin}{arg\,min}

\def\BibTeX{{\rm B\kern-.05em{\sc i\kern-.025em b}\kern-.08em
    T\kern-.1667em\lower.7ex\hbox{E}\kern-.125emX}}
\begin{document}

\title{Temporally Stable Multilayer Network Embeddings:\\ A Longitudinal Study of Russian Propaganda}

\DeclareRobustCommand{\IEEEauthorrefmark}[1]{\smash{\textsuperscript{\footnotesize #1}}}

\iftrue

\makeatletter
\newcommand{\linebreakand}{%
  \end{@IEEEauthorhalign}
  \hfill\mbox{}\par
  \mbox{}\hfill\begin{@IEEEauthorhalign}
}
\makeatother

\author{
    \IEEEauthorblockN{
        Daniel Matter\IEEEauthorrefmark{1}\orcidlink{0000-0003-4501-5612}, 
        Elizaveta Kuznetsova\IEEEauthorrefmark{2}\orcidlink{0000-0002-3614-1804},
        Victoria Vziatysheva\IEEEauthorrefmark{2}\orcidlink{0000-0002-3762-6758},
        Ilaria Vitulano\IEEEauthorrefmark{2},
        Jürgen Pfeffer\IEEEauthorrefmark{1}\orcidlink{0000-0002-1677-150X}
    }
    \\
    \linebreakand
    \IEEEauthorblockA{
        \IEEEauthorrefmark{1}
        \textit{School of Social Sciences and Technology} \\
        \textit{Technical University of Munich}\\
        Munich, Germany \\
        \{daniel.matter, juergen.pfeffer\}@tum.de 
    }
    \and
    \IEEEauthorblockA{
        \IEEEauthorrefmark{2}
        \textit{Weizenbaum Insititut}\\
        Berlin, Germany \\
        \{elizaveta.kuznetsova, victoria.vziatysheva, ilaria.vitulano\}\\@weizenbaum-institut.de 
    }
}
\fi

\maketitle

\begin{abstract}
Russian propaganda outlet RT (formerly, Russia Today) produces content in seven languages. There is ample evidence that RT's communication techniques differ for different language audiences. In this article, we offer the first comprehensive analysis of RT's multi-lingual article collection, analyzing all $2.4$ million articles available on the online platform from 2006 until 06/2023. Annual semantic networks are created from the co-occurrence of the articles' tags. Within one language, we use AlignedUMAP to get stable inter-temporal embeddings. 
Between languages, we propose a new method to align multiple, sparsely connected networks in an intermediate representation before projecting them into the final embedding space. With respect to RT's communication strategy, our findings hint at a lack of a coherent strategy in RT's targeting of audiences in different languages, evident through differences in tag usage, clustering patterns, and uneven shifts in the prioritization of themes within language versions. Although identified clusters of tags align with the key themes in Russian propaganda, such as Ukraine, foreign affairs, Western countries, and the Middle East, we have observed significant differences in the attention given to specific issues across languages that are rather reactive to the information environment than representing a cohesive approach.
\end{abstract}


\section{Introduction}
In recent years, digital information environments have undergone a profound transformation \cite{blumler_fourth_2016}. While changes like diversification of content and communication abundance have enabled positive public discourse dynamics in democratic societies, they have at the same time created new opportunities for problematic content dissemination \cite{blumler_fourth_2016}, \cite{howard_algorithms_2018}. Although propaganda, misinformation, and disinformation are not new phenomena, they have taken on new dimensions within algorithmically mediated information environments. Developments in automation and marketing techniques aimed at search engine optimization (SEO) have led to an increase in computational propaganda \cite{woolley_computational_2019}, \cite{bradshaw_disinformation_2019}. Scholarly research has been preoccupied with the problem since the 2016 Presidential Election in the US \cite{howard_algorithms_2018} and had given it a new wave of attention during the COVID-19 pandemic, when virus and vaccine-related disinformation became a severe public threat \cite{muhammed_disaster_2022}. While some digital platforms have taken the problem of misinformed publics seriously and attempted to reduce user exposure to it, the intensity and the volume of propaganda content across platforms, as well as the virality of its distribution, are still alarmingly high \cite{allcott_trends_2019}. 

The Russian government has played a significant role in shaping the information environment rife with propaganda and false narratives, particularly by its main propaganda arm RT (formerly, Russia Today) \cite{elswah_anything_2020}. Conceived in 2005 as an international broadcaster and part of a Kremlin-funded program to promote the image of Russia abroad, it has since become an agenda-setter outlet for Russian state propaganda. Heavily reliant on the spread of disinformation and conspiracy theories, RT's influence on public opinion worldwide is especially concerning amid Russia's war in Ukraine \cite{yablokov_conspiracy_2015}. Studies to date focus on either the channel's Russian \cite{yuskiv_media_2021} or English \cite{hoyle_portrait_2021}, \cite{crilley_russia_2022} version, with very few accounts of the French, German \cite{muller_right-wing_2022}, Arabic \cite{dajani_differentiated_2021}, Spanish, and Serbian. Cross-language studies are even more rare \cite{orttung_russia_2018}. Existing works on RT are usually limited to specific case studies and investigate its organizational structure and culture \cite{elswah_anything_2020}, \cite{kuznetsova_kontrpropaganda_2021} or its tools and methods of influence \cite{crilley_russia_2022}, \cite{glazunova_soft_2022}. However, we still lack a holistic understanding of RT's strategic output and the evolution of its narratives over time. Existing studies on misinformation, disinformation, and digital propaganda are disproportionally focused on social media \cite{altay_misinformation_2023} and rarely include analyses of source websites themselves. Moreover, large-scale analysis of Russian foreign propaganda is challenging to implement due to the lack of complete datasets that are comparable across different languages.

While existing studies primarily work with the social media content of RT, which uses the format defined by a platform, we present a comparative multilingual survey on its entire output across RT's language versions. To do so, we propose a new method to create multilayer network embeddings that adhere to a two-dimensional grid of constraints. We leverage AlignedUMAP to generate intermediate, temporally stable embeddings, which are then aligned between languages and projected into the final embeddings space. 

The main contributions of this article are:
\begin{itemize}
    \item Proposing a method to generate network embeddings for multilayered networks with a complex constraints structure.
    \item Showcasing how to create temporally stable clusters from these embeddings, which can be used to examine the genesis of topics within each language and between languages.
    \item A comprehensive network of tags and its longitudinal analysis highlight significant changes in the attention given to specific issues on different RT versions and a lack of a coherent approach in RT's targeting of audiences in different languages.
    \item Making the dataset of all $2.4$ million RT articles available to the scientific community.
\end{itemize}

\section{Related Work}
\textbf{RT and Russian propaganda.}
The Russian government is known to actively pollute the international online information environment with state narratives and disinformation targeted at domestic and foreign audiences \cite{helmus_russian_2018}. It deploys diverse tools to spread digital propaganda including the use of state-sponsored media, trolls, and bots\cite{vesselkov_russian_2020}, as well as extensive networks of social media accounts (e.g., on Telegram) \cite{vavryk_mapping_2022}. RT, perhaps the most prominent Russian broadcaster, plays a central role in defining the strategy of the Kremlin's propaganda \cite{helmus_russian_2018}. 
Existing research has found that RT strategically focuses on influencing multilingual audiences in different parts of the world, including Russia, the West, Latin America, and the Arab world, through targeted promotion of Russia's strategic narratives \cite{orttung_russia_2018}. Its methods of influence include dissemination of conspiracy theories\cite{yablokov_conspiracy_2015}, targeted use of humor and sarcasm \cite{crilley_russia_2020}, disinformation, and defensive rhetoric aimed at justifying Russia's actions \cite{kuznetsova_kontrpropaganda_2021}. Most studies to date focus on the English version of the outlet\cite{elswah_anything_2020} with only a few accounts performing cross-language studies. 

Researchers use varying data sets to study RT. Orttung and Nelson \cite{orttung_russia_2018} collected 70,220 videos from RT's Youtube channel. 
They performed qualitative text analysis of video titles to identify the target audience. They qualitatively categorized all videos into 'nine geographical' designations and compared the results with the number of views per category to identify the potential effect on the audiences.  Glazunova et al. \cite{glazunova_soft_2022} also investigate the sharing of RT's content in six languages. Focusing on Facebook communities, they collected a data set of 914,615 posts on the platform using CrowdTangle and constructed a network of sharing of RT content on Facebook, focusing on audiences. 

Unlike previous comparative studies focusing on mostly social media data and more static network analyses, we study changes over time utilizing a comprehensive dataset directly from RT's original platform. 


\textbf{Multimodal and temporal network data.} 
Network analysis has been a stable of social media, and more broadly internet-related research for multiple decades now \cite{borgattiNetworkAnalysisSocial2009, delfresnogarciaIdentifyingNewInfluences2016}, as exemplified by the analysis of propaganda networks done by  Pyo \cite{pyoNetworkPropagandaManipulation2019}.

The recent surge in multimodal network data, e.g. from social media, has spiked interest in network models capable of handling such data. 
Ghoniem et al. Bonifazi et al. \cite{bonifaziInvestigatingCOVID19Vaccine2022} used multilayer networks to model user groups on Twitter related to the Covid-19 Pandemic, while Oro et al. \cite{oroDetectingTopicAuthoritative2018} employed multilayer networks to detect topic authoritative social media users.
\cite{ghoniemStateArtMultilayer2019} gives a comprehensive overview of the state of the art in multilayer network visualization.

Modeling time in complex networks has recently seen particular research interest \cite{salamaTemporalNetworksReview2022}.
While De Domenico et al. \cite{dedomenicoMathematicalFormulationMultilayer2013} formalized central concepts from single-layer networks for multilayer networks, 
Bazzi et al. \cite{bazziCommunityDetectionTemporal2016} and \cite{ghawiCommunityMatchingBased2022} demonstrated methods for community detection in multilayer networks with a time component.



\section{Methodology}
\subsection{Data}
Russia Today has online outlets in seven languages\footnote{Arabic (AR), English (EN), French (FR), German (DE), Spanish (ES), Russian (RU), Serbian (RS)}. 
Leveraging the pages' sitemaps, we scraped all articles from all of these pages. Of all 2'468'897 unique sites, 378 were unreachable, which we suspect to be permanently deleted articles. Information about the remaining articles can be seen in Table \ref{table:descr}.

Due to recent blockages by the European Union, some of these pages exist under multiple URLs, e.g., the German version is hosted at \url{https://de.rt.com/}, which is not reachable within the EU, as well as at \url{https://pressefreiheit.rtde.live/}, which is reachable globally---we can confirm that both versions host the same content. 

\begin{table}[htbp]
\caption{Dataset Descriptives}
\label{table:descr}
\begin{center}
\begin{tabular}{|c|c|r|r|r|}
\hline
\textbf{Lang.} & \textbf{\textit{First Article}}& \textbf{\textit{Nr. Articles}} & \textbf{\textit{Nr. Tags}} & \textbf{\textit{Nr. Authors}} \\
\hline
AR & 2005-01-01 & 710,529 & 30 & N/A \\
DE & 2014-09-03 & 90,484 & 2,449 & N/A \\
EN & 2006-06-21 & 257,546 & 894 & 140 \\
ES & 2009-01-07 & 407,670 & 927 & 30 \\
FR & 2015-01-12 & 68,502 & N/A & N/A \\
RU & 2008-06-23 & 909,572 & 3,186 & 588 \\
RS & 2022-09-22 & 24,216 & 2,022 & 79 \\
\hline
\end{tabular}
\end{center}
\end{table}

\subsection{Network Construction and Analysis}
We use the tags assigned to each article to analyze the semantic output distribution of each outlet over time.
In particular, we construct one network per language per year, where we consider each tag as a node, and the concurrence of two tags in the same article within each year as the edge weight between those nodes in that year.
As tag usage differs in different languages, we restrict each network to only contain the 200 most-used tags per year.

We consider data from 2018 until 2023, as most outlets started adopting proper tag usage around or before 2018.
Disregarding Arabic and French for their lack of tags, and Serbian, as it was only recently established, we are left with the English, German, Spanish, and Russian versions of RT. Over six years, this yields a total of 24 individual networks.
Understanding the evolution of these networks depends on the ability to generate stable embeddings over time and between languages. 
Figure \ref{fig:flow-schema} describes our approach given these intertemporal and interlingual links.

In step (I), within each language, the embedding position of the same tag should be as stable as possible across years. Fixing it completely, however, is not sensible. Should the surrounding of a tag change vastly, i.e., if there is a shift in how it is used, we want the tag to be able to move around within the network.
In step (II), we want to embed similar tags into close coordinates between languages. After a clustering step, we use a multilingual BERT model \cite{devlinBERTPretrainingDeep2019} to find semantically related regions, i.e., clusters, between networks of different languages from the same year. Only Russian, possibly due to the non-Latin script, needed manual intervention, as it did not produce sensible pairings. 
Lastly, in step (III), we transform all intermediate embeddings into the final, two-dimensional embeddings space.

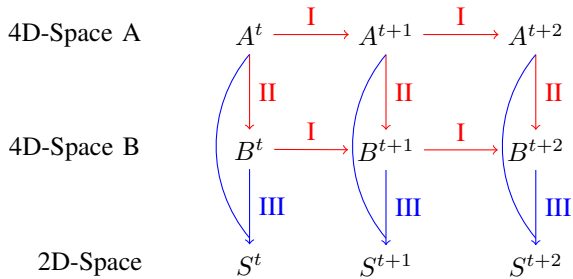
\begin{figure}[b]
    \centering
    \begin{tikzpicture}
    
        \node[](nat1)[]{$A^t$};
        \node[](nat2)[right=of nat1]{$A^{t+1}$};
        \node[](nat3)[right=of nat2]{$A^{t+2}$};

        \node[](nbt1)[below=of nat1]{$B^t$};
        \node[](nbt2)[below=of nat2]{$B^{t+1}$};
        \node[](nbt3)[below=of nat3]{$B^{t+2}$};

        \node[](st1)[below=of nbt1]{$S^t$};
        \node[](st2)[below=of nbt2]{$S^{t+1}$};
        \node[](st3)[below=of nbt3]{$S^{t+2}$};

        \node[align=center](lA)[left=of nat1]{4D-Space A};
        \node[align=center](lB)[left=of nbt1]{4D-Space B};
        \node[align=center](s)[left=of st1]{2D-Space};

        \draw[->, color=red] (nat1) -- node[above]{I} (nat2);
        \draw[->, color=red] (nat2) -- node[above]{I} (nat3);

        \draw[->, color=red] (nbt1) -- node[above]{I} (nbt2);
        \draw[->, color=red] (nbt2) -- node[above]{I} (nbt3);

        \draw[->, color=red] (nat1) -- node[right]{II} (nbt1);
        \draw[->, color=red] (nat2) -- node[right]{II} (nbt2);
        \draw[->, color=red] (nat3) -- node[right]{II} (nbt3);
        
        \draw[->, color=blue] (nbt1) -- node[right]{III} (st1);
        \draw[color=blue] (nat1.south) arc[radius=1.9, start angle=140, end angle=220] (st1.north);
        \draw[->, color=blue] (nbt2) -- node[right]{III} (st2);
        \draw[color=blue] (nat2.south) arc[radius=1.9, start angle=140, end angle=220] (st2.north);
        \draw[->, color=blue] (nbt3) -- node[right]{III} (st3);
        \draw[color=blue] (nat3.south) arc[radius=1.9, start angle=140, end angle=220] (st3.north);

    \end{tikzpicture}
    \caption{Algorithm Flow-Chart. Optimal-Alignment-Constraints are red, and dimensionality-reduction is blue.}
    \label{fig:flow-schema}
\end{figure}

\subsubsection{Intertemporal Alignment}
Intertemporal alignment within each language can be posed as a simple multilayer network problem.
AlignedUMAP \cite{mcinnesUMAPUniformManifold2020} provides an adequate way to generate embeddings from a list of sequentially chained networks, as has been demonstrated by others \cite{daduApplicationAlignedUMAPLongitudinal2023, rahimiANTMAlignedNeural2023}. It allows us to map nodes from the current layer to nodes from the previous one. Connections across multiple layers are not possible. However, since around $90\%$ of nodes remain stable from one year to the next, this should not strongly affect the overall performance. 

We chose cosine-similarity as the input metric and output into a four-dimensional space with the standard L2-norm. The choice of output dimension and metric, i.e., the space into which the intermediate representation is embedded, is crucial.

A high-dimensional intermediate representation yields much better performance in the interlingual alignment step, as the number of tuneable parameters grows quadratically in the dimension, and we restrict ourselves to linear transformations.
On the other hand, a low-dimensional intermediate representation ensures that a lot of structure is captured within each language, which might be lost in the third dimensionality-reduction step. 
For this study, we chose $D = 4$, as it provided the best trade-off between those contending goals.
Choosing the L2-norm for the intermediate representation is sensible, as it allows us to formulate the interlingual alignment problem in terms of an ordinary least-squares approximation.

\subsubsection{Clustering}
Clustering is critical for a robust understanding of the content at hand. We perform HDBSCAN \cite{mcinnesHdbscanHierarchicalDensity2017} on each of the 24 intermediate representations. Thanks to the intertemporal alignment, which minimizes changes in embedding positions and hence network density as far as possible, this yields very stable clusters across time. 
We use these clusters twofold. Firstly, in section \ref{sec:res}, we analyze their prevalence shifts in content over time. 
Secondly, similar to \cite{ghawiCommunityMatchingBased2022}, we use them to achieve robust interlingual links. For each cluster, we compute BERT-Embeddings for the three most prevalent terms, average them, and look for similar clusters in all other languages, using cosine distance as a metric. If we find a cluster above a set similarity threshold\footnote{We set this threshold such that the interlingual similarity has to be at least as good as the best matching cluster within the pair of languages.}, we use the most similar term pair and consider them an interlingual link.

\subsubsection{Interlingual Alignment}
Given the intermediate representation and interlingual links from the previous step, we now align the networks of different languages. While it is possible to align all languages simultaneously, we decided to keep one language, English, fixed and align all other languages against it,
since RT was initially mainly focused on English-speaking audiences, and to this date, many articles in other languages are direct translations from English.

If the intermediate representation of two languages from the same year is given by $A^t \in \mathbb{R}^{N_1 \times D}, B^t \in \mathbb{R}^{N_2 \times D}$, $\hat{A}^t, \hat{B}^t \in \mathbb{R}^{M \times D}$ shall denote those sub-graphs that contain only the tags which are linked between those languages. Hence, we want to bring all points $\hat{A}_i^t, \hat{B}_i^t$ as close to each other as possible. Formally, we compute

\begin{align*}
    P_{A, B}^t &= \argmin_{X} || \begin{pmatrix} & \dots \\ \hat{A}^t & \mathbf{1} \\ & \dots \end{pmatrix} X - \hat{B}^t ||_2
\end{align*}
\begin{align*}
    \mathbf{A}^t &= \begin{pmatrix} & \dots \\ A^t & \mathbf{1} \\ & \dots \end{pmatrix} P_{A, B}^t
\end{align*}

We augment $\hat{A}^t$ with a column of ones to also allow for spatial translation of the embeddings and then use $P_{A, B}^t$ to apply the optimal projection to the entire network $A^t$, yielding $\mathbf{A}^t$, which is optimally aligned to $B$.
In our case, $B$ will always be English, the fixed background.

\subsubsection{Dimensionality Reduction}

Once aligned, we perform PCA to project from the four-dimensional intermediate representations into the final two-dimensional space. A single PCA is fitted on the entire dataset, i.e., all languages and all years at once, to prevent skewed results while preserving as much overall information as possible. More intricate dimensionality-reduction methods might be interesting if one uses higher-dimensional intermediate representations. However, in our case, PCA yields fast and robust results.

\begin{figure*}
    \centering
    \includegraphics[width=.925\linewidth]{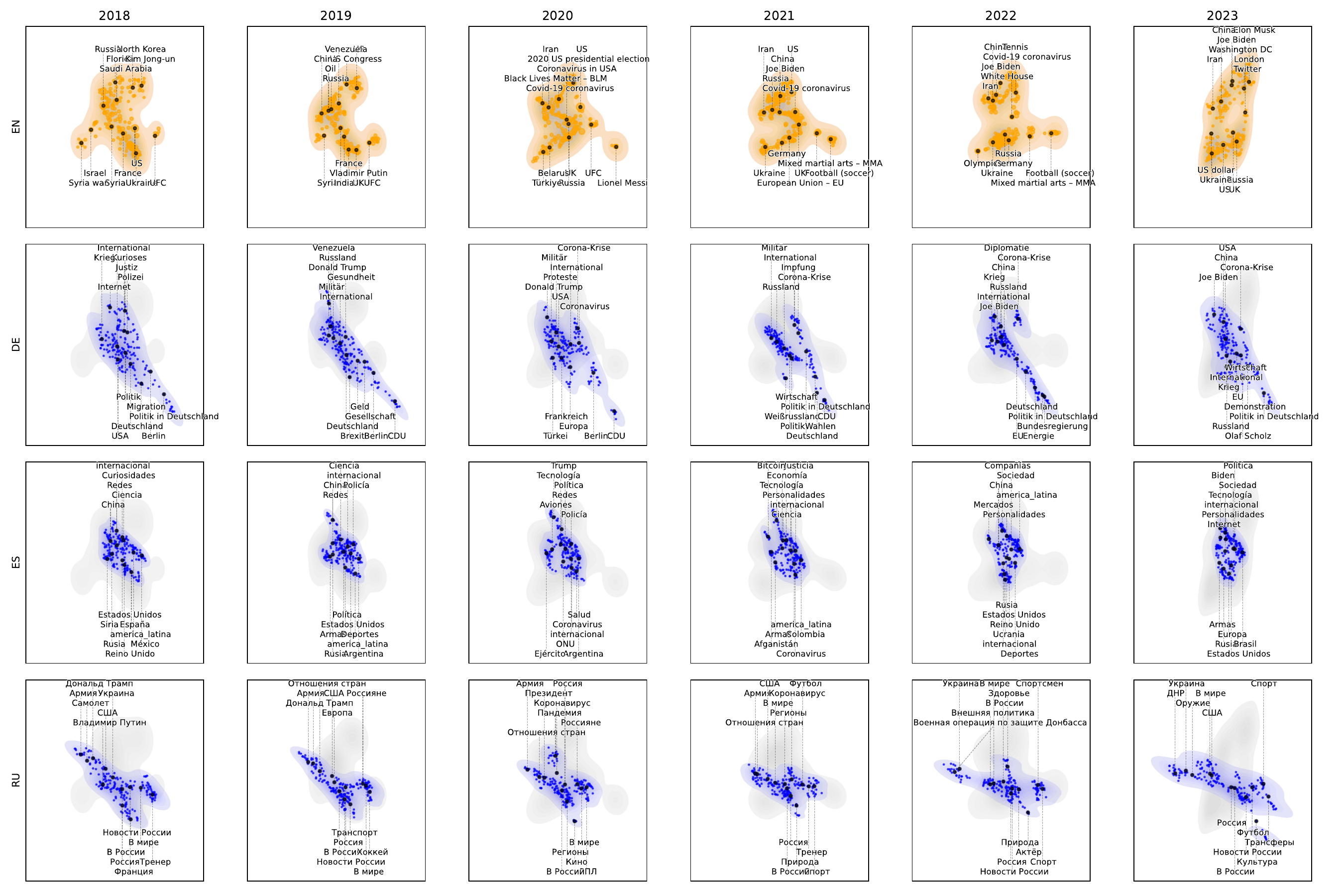}
    \caption{Density-Kernel of Final Embeddings. The upmost row only displays the English embeddings. In all other rows, the gray area in the background indicates the outline of the baseline---i.e., English---network.}
    \label{fig:overview}
\end{figure*}

Applying this pipeline to our data results in Figure \ref{fig:overview}.
Figure \ref{fig:clusters:prev} depicts the share of articles that contain a specific tag, the mean relative frequency of the terms within selected and partially joined clusters from step II. Note that the absolute number might be skewed due to concurrences. However, the trends are unaffected by this.
Figure \ref{fig:clusters:cohesion} displays the log-odds-ratio of observed versus expected tag occurrence within the same clusters. If $\alpha$ is the number of occurrences where both tags are within a cluster, $\beta$ is the number of occurrences with at least one tag in that cluster, the cluster has size $k$, and the network has size $N$, we calculate this as $\log \frac{\alpha}{\beta} - \log \frac{k(k-1)}{k(N-1)}$. This number effectively represents how much more likely a concurrence appears within a cluster than expected by chance.

\section{Results}
\label{sec:res}

\begin{table}[htbp]
\setlength\tabcolsep{1.5pt}
\caption{First three most prevalent tags per cluster}
\label{table:descr}
\begin{center}
\begin{tabular}{|p{1.5em}|p{29em}|}
\hline
\textbf{}&\textbf{Three most prevalent tags per Cluster}\\
\hline
\textbf{de}&\tiny \textbf{0}: Olympia,  Doping,  IOC; \textbf{1}: Sport,  Fußball,  Fußball-WM 2018; \textbf{2}: Serbien,  Balkan,  Kosovo; \textbf{3}: Südafrikanische Republik,  Mali,  Sudan; \textbf{4}: Brexit,  London,  Theresa May; \textbf{5}: Lateinamerika,  Venezuela,  Brasilien; \textbf{6}: Umwelt,  Klimawandel,  Natur; \textbf{7}: Kultur,  Kunst,  Musik; \textbf{8}: Verkehrswesen,  Autos,  Tesla; \textbf{9}: Brandenburg,  Deutsche Bahn,  Öffentliche; \textbf{10}: Schule,  Bildung,  Kinderrechte; \textbf{11}: Wissenschaft,  Forschung,  Weltall; \textbf{12}: CDU,  SPD,  Bündnis 90/Die Grünen; \textbf{13}: Berlin,  Bayern,  Sachsen; \textbf{14}: Syrien,  Türkei,  Terror; \textbf{15}: Frauenrechte,  LGBT,  Frauen; \textbf{16}: Sanktionen,  Energie,  Öl; \textbf{17}: Geld,  Armut,  Arbeitsmarkt; \textbf{18}: Handel,  Finanzsystem,  Handelskrieg; \textbf{19}: Armenien,  Aserbaidschan,  Kasachstan; \textbf{20}: Corona-Krise,  Coronavirus,  Gesundheit; \textbf{21}: Impfung,  Impfstoff,  Lockdown; \textbf{22}: Internet,  Soziale Medien,  Twitter; \textbf{23}: Asien,  Indien,  Japan; \textbf{24}: China,  Donald Trump,  Iran; \textbf{25}: Israel,  Saudi-Arabien,  Jemen; \textbf{26}: Nahost,  Palästina,  Gaza; \textbf{27}: Zweiter Weltkrieg,  Nazismus,  Sowjetunion; \textbf{28}: Medienkritik,  Propaganda,  Journalismus; \textbf{29}: BILD,  ZDF,  Der Spiegel; \textbf{30}: Wirtschaft,  Europa,  Interviews; \textbf{31}: Technik,  Technologie,  IT; \textbf{32}: Migration,  Flüchtlinge,  Flüchtlingskrise; \textbf{33}: Justiz,  Polizei,  Kriminalität; \textbf{34}: Emmanuel Macron,  Religion,  Islam; \textbf{35}: Politik,  EU,  Proteste; \textbf{36}: Geschichte,  Weißrussland,  Alexander Lukaschenko; \textbf{37}: International,  Russland,  Diplomatie; \textbf{38}: Militär,  Waffen,  Armee; \textbf{39}: Donbass,  Wladimir Selenskij,  Kriegsverbrechen; \textbf{40}: Moskau,  FSB,  Sankt Petersburg; \textbf{41}: Ukraine-Konflikt,  Wladimir Putin,  Sergei Lawrow\\ \hline 
\textbf{en}&\tiny \textbf{0}: Football (soccer),  Cristiano Ronaldo,  Lionel Messi; \textbf{1}: Iran tension,  Middle East,  US sanctions on Iran; \textbf{2}: Israel,  State of Palestine,  Gaza Strip; \textbf{3}: Brexit,  Coronavirus in UK,  Boris Johnson; \textbf{4}: Kurds in Syria,  Idlib,  Chemical weapon; \textbf{5}: Olympics,  Russian figure skating,  Fukushima disaster; \textbf{6}: Pakistan,  India and Pakistan relations,  Narendra Modi; \textbf{7}: Julian Assange,  WikiLeaks,  Ecuador; \textbf{8}: Emmanuel Macron,  Paris,  Coronavirus in France; \textbf{9}: President of Russia,  Russian presidential election 2018,  Putin presser; \textbf{10}: Recep Tayyip Erdogan,  Greece,  Türkiye-US relations; \textbf{11}: Syria war,  Islamic State of Iraq and the Levant – ISIS,  Terrorism; \textbf{12}: Coronavirus in USA,  Coronavirus vaccines,  Tennis; \textbf{13}: Covid-19 coronavirus,  Hungary,  Censorship; \textbf{14}: Coronavirus in Germany,  Angela Merkel,  Berlin; \textbf{15}: US,  Russia,  Ukraine; \textbf{16}: China-US relations,  Japan,  Asia; \textbf{17}: China,  Russia-US relations,  Nuclear weapons; \textbf{18}: Donald Trump,  Joe Biden,  2020 US presidential election; \textbf{19}: Faceboo,  Elon Musk,  Google; \textbf{20}: National Football League – NFL,  National Basketball Association – NBA,  Michigan; \textbf{21}: Iran,  Pentagon,  Afghanistan; \textbf{22}: LGBTQ,  Amazon,  Jeffrey Epstein; \textbf{23}: Racism,  Black Lives Matter – BLM,  Police brutality; \textbf{24}: New York city,  California,  Texas\\ \hline 
\textbf{es}&\tiny \textbf{0}: M,  A,  S; \textbf{1}: Deportes,  Fútbol,  Bolsonaro; \textbf{2}: Narcotráfico en México,  López-Obrador,  Paola Guzmán; \textbf{3}: Maduro,  Chávez,  Érika Sanoja; \textbf{4}: Ignacio Jubilla,  Ricardo Romero,  Macri; \textbf{5}: México,  Argentina,  Venezuela; \textbf{6}: Cataluña,  francisco-guaita,  Partidos; \textbf{7}: Redes,  Internet,  Compañías; \textbf{8}: Salud,  Coronavirus,  Tecnología; \textbf{9}: Segunda Guerra Mundial,  URSS,  Gran Guerra Patria; \textbf{10}: Moscú,  Siberia,  San Petersburgo; \textbf{11}: Espacio,  Descubrimientos,  Astronomía; \textbf{12}: Nigeria,  Somalia,  Kenia; \textbf{13}: Rock,  Concierto,  Pop; \textbf{14}: Petróleo,  Crisis económica,  Mercados; \textbf{15}: Accidentes,  Desastres naturales,  Asia; \textbf{16}: Ciencia,  Curiosidades,  Animales; \textbf{17}: Cultura,  Cine,  Música; \textbf{18}: internacional,  Trump,  Obama; \textbf{19}: Policía,  Delincuencia,  Justicia; \textbf{20}: Armas,  Ejército,  Defensa; \textbf{21}: Política,  Rusia,  Economía; \textbf{22}: Unión Europea,  UE,  Grecia; \textbf{23}: España,  Reino Unido,  Europa; \textbf{24}: Islam,  Libia,  Yemen; \textbf{25}: Israel,  Arabia Saudita,  Programa nuclear de Irán; \textbf{26}: Terrorismo,  Siria,  Conflicto en Siria\\ \hline 
\textbf{ru}&\fontsize{5}{4}\selectfont \textbf{0}: \textcyrillic{Сборная России по биатлону}, \textcyrillic{Кубок мира по биатлону}, \textcyrillic{СБР}; \textbf{1}: \textcyrillic{Австралия}, \textcyrillic{Турниры Большого шлема}, \textcyrillic{Даниил Медведев}; \textbf{2}: \textcyrillic{Смешанные единоборства}, \textcyrillic{UFC}, \textcyrillic{Хабиб Нурмагомедов}; \textbf{3}: \textcyrillic{НХЛ}, \textcyrillic{КХЛ}, \textcyrillic{Сборная России по хоккею}; \textbf{4}: \textcyrillic{Александр Овечкин}, \textcyrillic{Кубок Стэнли}, \textcyrillic{Никита Кучеров}; \textbf{5}: \textcyrillic{Евгения Медведева}, \textcyrillic{Алина Загитова}, \textcyrillic{Александра Трусова}; \textbf{6}: \textcyrillic{Сюжеты RT}, \textcyrillic{LIVE видео}, \textcyrillic{Происшествия}; \textbf{7}: \textcyrillic{Лондон}, \textcyrillic{Брексит}, \textcyrillic{Сергей Скрипаль}; \textbf{8}: \textcyrillic{Наука}, \textcyrillic{Религия}, \textcyrillic{Учёные}; \textbf{9}: \textcyrillic{Кремль}, \textcyrillic{Указ}, \textcyrillic{Пресс-конференция Владимира Путина для СМИ}; \textbf{10}: \textcyrillic{Самолет}, \textcyrillic{Авиация}, \textcyrillic{Авиакатастрофа}; \textbf{11}: \textcyrillic{В мире}, \textcyrillic{Россияне}, \textcyrillic{Франция}; \textbf{12}: \textcyrillic{Коронавирус}, \textcyrillic{Здоровье}, \textcyrillic{Пандемия}; \textbf{13}: \textcyrillic{Олег Табаков}, \textcyrillic{Николай Караченцов}, \textcyrillic{Дмитрий Марьянов}; \textbf{14}: \textcyrillic{История}, \textcyrillic{Великая Отечественная война}, \textcyrillic{СССР}; \textbf{15}: \textcyrillic{Хоккей}, \textcyrillic{Фигурное катание}, \textcyrillic{Теннис}; \textbf{16}: \textcyrillic{Допинг}, \textcyrillic{Лёгкая атлетика}, \textcyrillic{МОК}; \textbf{17}: \textcyrillic{Спортсмен}, \textcyrillic{Тренер}, \textcyrillic{Трансферы}; \textbf{18}: \textcyrillic{РПЛ}, \textcyrillic{ФК Спартак}, \textcyrillic{ФК Зенит}; \textbf{19}: \textcyrillic{Артём Дзюба}, \textcyrillic{Сергей Семак}, \textcyrillic{Фёдор Смолов}; \textbf{20}: \textcyrillic{Культура}, \textcyrillic{Кино}, \textcyrillic{Искусство}; \textbf{21}: \textcyrillic{Конкурс}, \textcyrillic{Песня}, \textcyrillic{Концерт}; \textbf{22}: \textcyrillic{Военная операция по защите Донбасса}, \textcyrillic{Донбасс}, \textcyrillic{ДНР}; \textbf{23}: \textcyrillic{Киев}, \textcyrillic{Пётр Порошенко}, \textcyrillic{Владимир Зеленский}; \textbf{24}: \textcyrillic{Видео}, \textcyrillic{Армия}, \textcyrillic{Минобороны}; \textbf{25}: \textcyrillic{Смерть}, \textcyrillic{СМИ}, \textcyrillic{Полиция}; \textbf{26}: \textcyrillic{Сирия}, \textcyrillic{Терроризм}, \textcyrillic{Исламское государство}; \textbf{27}: \textcyrillic{Уголовное дело}, \textcyrillic{Прокуратура}, \textcyrillic{Сочи}; \textbf{28}: \textcyrillic{Иран}, \textcyrillic{Азия}, \textcyrillic{Южная Корея}; \textbf{29}: \textcyrillic{Турция}, \textcyrillic{Взрыв}, \textcyrillic{Теракт}; \textbf{30}: \textcyrillic{Германия}, \textcyrillic{ЕС}, \textcyrillic{Великобритания}; \textbf{31}: \textcyrillic{США}, \textcyrillic{Отношения стран}, \textcyrillic{МИД}; \textbf{32}: \textcyrillic{Дональд Трамп}, \textcyrillic{Барак Обама}, \textcyrillic{Нью-Йорк}; \textbf{33}: \textcyrillic{Регионы}, \textcyrillic{Автомобиль}, \textcyrillic{Транспорт}; \textbf{34}: \textcyrillic{Госдеп США}, \textcyrillic{Джо Байден}, \textcyrillic{Белый дом}; \textbf{35}: \textcyrillic{Президентские выборы в США}, \textcyrillic{Хиллари Клинтон}, \textcyrillic{Демократическая партия}; \textbf{36}: \textcyrillic{Экономика}, \textcyrillic{Энергетика}, \textcyrillic{Газ}; \textbf{37}: \textcyrillic{Деньги}, \textcyrillic{Финансы}, \textcyrillic{Бизнес}; \textbf{38}: \textcyrillic{РИА Новости}, \textcyrillic{Эксперт}, \textcyrillic{Крым}; \textbf{39}: \textcyrillic{В России}, \textcyrillic{Дети}, \textcyrillic{Образование}; \textbf{40}: \textcyrillic{Новости России}, \textcyrillic{ЧП}, \textcyrillic{Пожар}\\ \hline 

\hline
\end{tabular}
\end{center}
\end{table}

Looking at Figure \ref{fig:overview}, we can observe the evolution of tags over time within one language and compare them between languages. While there is considerable overlap, we can see that the German and Russian versions, particularly, have vast areas that English does not cover. Further, both COVID-19, as well as the invasion of Ukraine change the overall shape of the networks, i.e., the content distribution.

Going into more detail, we identified the main clusters of tags for each language version (see Table II). Typically, tags in our dataset represent geographic locations (e.g., “Ukraine”, “Europe”, “New York”), names of people (e.g., “Sergey Lavrov”, “Joe Biden”, “Vladimir Putin”), and organizations (e.g., “World Bank”, “Der Spiegel”, “RT”), or topic-specific terms (e.g., “Lockdown”, “SWIFT”, “Russophobia”).

\begin{figure*}[ht]
    \centering
    \includegraphics[width=\linewidth]{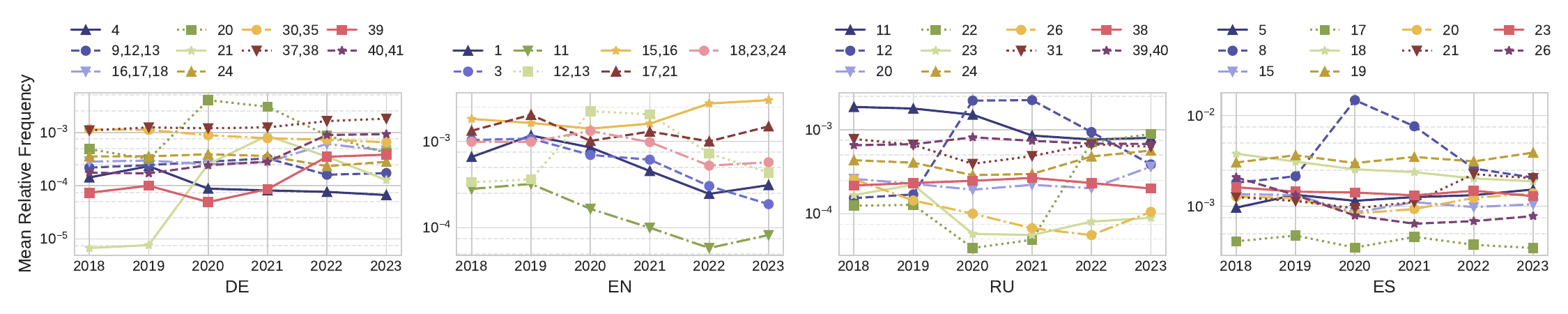}
    \caption{Mean frequency of articles containing the tags within each cluster for all analyzed languages.}
    \label{fig:clusters:prev}
\end{figure*}

\begin{figure*}[ht]
    \centering
    \includegraphics[width=\linewidth]{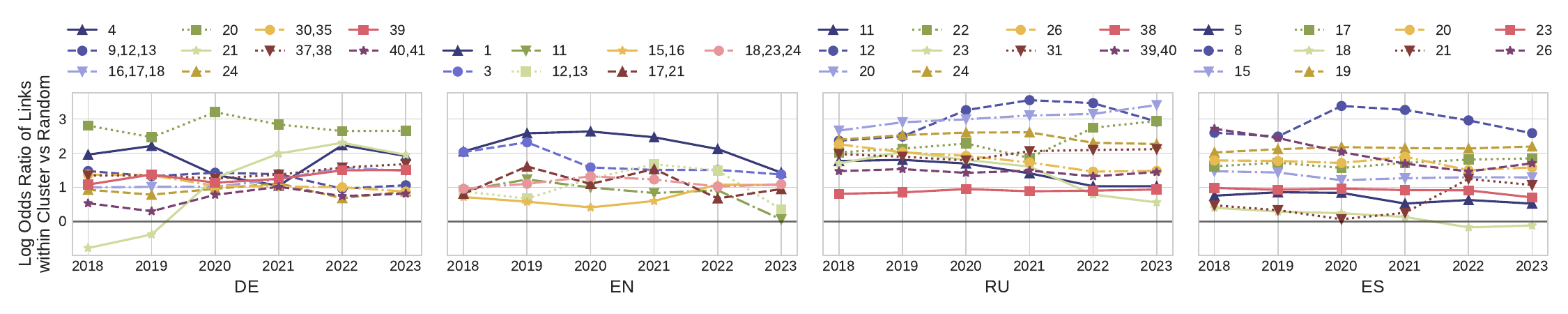}
    \caption{Ratio of the probability of a tag occurring next to another tag of the same cluster against the probability of random cooccurrences, i.e., cluster cohesiveness.}
    \label{fig:clusters:cohesion}
\end{figure*}

The four language versions analyzed in the study use tags with different frequencies. Importantly, different RT language versions employ distinct strategies of tag usage. For example, RT in Russian uses more tags than any version and has a more fine-grained semantic matching of topics and a better-defined thematic clustering, while RT in English uses tags with a broader meaning. On the one hand, this can be explained by the fact that RT has different communication goals and uses different methods of targeting international audiences \cite{elswah_anything_2020}. On the other hand, it can be due to differences in journalistic practices and cultures \cite{kuznetsova_kontrpropaganda_2021}. Moreover, we have noticed inconsistencies in how tags are assigned to articles even within the same language version (e.g., tags with the same meaning may appear more than once if, for instance, the term is spelled differently or a name is used with or without a surname). Thus, we compared only the general trends that can be observed across different versions of RT. For representation purposes, we merged some distinct clusters that are thematically close to each other based on our qualitative analysis (see Figure \ref{fig:clusters:cohesion}).
First, some topics appear in multiple language versions, as shown by Figure \ref{fig:overview}. For instance, clusters related to Russia, Ukraine, COVID-19, and the Middle East appear in all languages. However, in some versions, they may be more extensive and detailed than in others. In Spanish, COVID-19 appears as a part of a general health-related cluster, while in German, Russian, and English, COVID clusters are among the most cohesive ones, as shown in Figure \ref{fig:clusters:cohesion}. In other versions, they may be represented by several clusters highlighting different topic dimensions.
Second, each version has prominent clusters related to the region which they may target: the US (en-18, en-23, en-24) and the UK (en-3), Russia (ru-38, ru-39, ru-40), Germany (de-12), Latin America (es-5). 
In the most prominent clusters, we can observe similar trends for the COVID-19-related clusters (spike in 2020-2021) and Ukraine-related clusters (spike in 2022-2023). 

\subsection{RT in Russian}

The Russian version of RT contains the largest network of tags (3,186), split into 41 clusters. RT was conceived as an international broadcaster aimed at foreign audiences. Its Russian language version was a late addition to the English, Spanish, and Arabic versions. Therefore, similar to its foreign audiences, Russian RT focused more on international affairs (ru-11) than Russian domestic topics (ru-39) in 2018. In 2020 we notice a strong reshuffling of this setting, primarily due to the spike of the COVID-19 pandemic cluster (ru-12): While the international affairs cluster (ru-11) consistently goes down between 2018 and 2022, the Russian domestic affairs cluster (ru-39) grows, indicating a strategic shift in focus. Interestingly, the international affairs cluster does not re-emerge after the pandemic focus subsides (ru-11) in 2022 but rather remains at a consistent level. Amid the war in Ukraine, the international relations cluster (ru-31) that mainly focuses on NATO-related tags receives a boost of attention in 2022. We identified two distinct clusters related to the topic of Ukraine: ru-22 is focused on the Eastern Ukrainian regions (\textcyrillic{"Donetsk and Luhansk"}) and the war events, including tags like "Ukrainian Armed Forces" (\textcyrillic{"ВСУ"}), "Military operation" (\textcyrillic{"Военная операция"}), "Kherson region" (\textcyrillic{"Херсонская область"}); ru-23 consists of tags related to Ukraine politics  (\enquote{Vladimir Zelenskyy}, \textcyrillic{"Владимир Зеленский"}; \enquote{Verkhovna Rada}, \textcyrillic{"Верховная рада"}; “Nationalism”, \textcyrillic{"Национализм"}; and \enquote{Maidan}, \textcyrillic{"Майдан"}). The two clusters follow similar patterns between 2018 and 2021, but the first one (ru-22) spikes in 2022. Overall, Ukraine-related tags form well-defined clusters on Russian RT since 2018, which indicates this topic was of strategic importance for RT in Russian even before the war. 

\subsection{RT in English}

The English version of RT uses fewer tags than the other three versions (894), making up a network of 25 distinct clusters. Similarly to all other language versions, RT in English has a spike in coronavirus-related tags (en-12) in 2020 and is consistently covering US domestic issues (en-18, en-23, en-24) and the Middle East (en-1) throughout the years. Given that this version of RT is targeted at English-speaking audiences, it has several US-related clusters(en-18, en-23, en-24) and a UK-focused cluster of tags (en-3). While the US clusters (en-18, en-23, en-24) have been consistently active throughout the years, the UK cluster has steadily declined since 2018. Unlike other languages, RT in English treats Ukraine as part of the international affairs theme (en-15). It remains consistently prevalent throughout the years and spikes in 2022 amid Russia’s war in Ukraine, while the US domestic affairs cluster goes down in that period (en-18, en-23, en-24). Particular attention on RT in English is given to terrorism (en-11) and the US relationship with non-western countries (en-17), containing tags like “Russia-US relations”, “US-backed coup in Venezuela”, and “Cold War”. The latter cluster spikes at the same time as the Ukraine-related coverage in 2022 and 2023, which may be connected to RT focusing on stories that support narratives about US expansionist and imperialist behavior \cite{yablokov_conspiracy_2015}.

\subsection{RT in German}

In the German version of RT, the network of 2,449 tags was divided into 42 distinct clusters. They relate to a variety of political, social, and cultural issues and focus on different regions of the world, e.g., the Balkans (de-2), the UK (de-4), and Latin America (de-5). Some clusters are more cohesive and isolated from the rest of the network (e.g., de-20, COVID-19, compare figure \ref{fig:clusters:cohesion}). Others are located more closely and have more overlaps (for example, clusters related to Europe). 
We observe considerable spikes for coronavirus-focused clusters: for coronavirus-related tags in general (de-20) in 2020 and for tags related to vaccines and pandemic health policies (de-21) in 2021. By 2023, both have dropped significantly. 
Other prominent clusters in the German version are related to international relations and the military (de-37 and de-38), as well as to the EU (de-30 and de-35). The first group of clusters contains tags like "NATO", "Geopolitics"("Geopolitik"), "Militarisation" ("Militarisierung") whereas the second---ones like "Europe" ("Europa"), "Election campaign" ("Wahlkampf") and multiple countries' names. In 2018, tags belonging to these clusters appeared in a similar number of articles. Yet, they show opposite dynamics over the years with the focus shifting towards international relations and military and away from the EU. The spike in the former is especially evident in 2022. After the beginning of Russia’s full-scale invasion of Ukraine, we also observe spikes in Ukraine- and Russia-focused clusters (de-39, de-40, and de-41) and those related to economics and finance (de-16, de-17, and de-18). The latter includes both general terms and those potentially related to sanctions against Russia such as "Sanctions (“Sanktionen”), "Energy crisis" (“Energiekrise”), and “Inflation”. The clusters of tags on German politics and domestic affairs (de-9, de-12, de-13) slightly increases by the beginning of the pandemic and peaked in 2021, which is likely related to the German elections. However, afterward, their activity drops again. 

\subsection{RT in Spanish}

The Spanish version of RT uses 927 tags, making up a network of 27 clusters. Similarly to other languages, the Spanish version has distinct clusters focusing on international affairs (es-18) and the middle east (es-26). The share of articles with tags from these two clusters steadily decreases between 2018 and 2023, while topics related to protests and justice (es-19) remain stable, indicating a shift in focus from international politics to domestic and societal issues. The cluster contains tags such as “Police” (“Policía”), “Justice” (“Justicia”), and “Protests” (“Protestas”). Articles referring to this cluster comprise a significant share of all the articles. The Health cluster (es-8) performs consistently with the other RT versions and spikes in 2020 due to the COVID-19 pandemic outbreak. Unlike the other languages, however, COVID does not constitute a cluster but is a single tag within this cluster (the second most used tag), indicating a less fine-grained use of tags for the topic. For the Spanish language, we identified two clusters related to the war in Ukraine. The first cluster was related to military topics in general (es-20), containing tags such as “Weapons” (“Armas”), “Defense” (“Defensa”), and “Nuclear Weapons” (“Armas Nucleares”). The second cluster focused more specifically on Ukraine (es-21) with tags such as “Russia” (“Rusia”), “Ukraine” (“Ucrania”), and “Conflicts” (“Conflictos”). Both clusters, located mainly at the center of the network, increased their shares in 2022, due to Russia's full-scale invasion of Ukraine.

\subsection{Summary}

Across all language versions, we find clusters somewhat associated with themes critical for Russian propaganda, such as Ukraine \cite{hutchings_dominant_2015}, international relations, Western countries\cite{hoyle_portrait_2021}, and the Middle East \cite{crilley_emotions_2020}. In all four language versions, we can trace a shift in RT's focus, first, during the pandemic and, second, after the beginning of Russia’s war against Ukraine. We also observe less expected structural changes in tags used by RT. For example, we see how the local agenda is being overridden by the international agenda in the context of the war or how local news become more dominant during COVID. Such trends can be observed on RT in German and Russian, respectively. Our findings also highlight how some issues represented by tags gain different levels and types of attention across languages: While health and pandemic-related tags form distinct clusters in the English and the Russian versions, on RT in Spanish, the cluster shows much less nuance and in Germany, they are split in two. The Ukraine topic is presented with a higher level of granularity on RT in Russian, forming two distinct clusters with separate foci and maintaining a consistent presence throughout the years, while RT in English does not treat it as a standalone theme.

\section{Conclusion}
\textbf{Contribution.}
We propose a novel approach to solving interlocked multilayer network layout problems. By leveraging an intermediate representation, we demonstrate how to align nodes in different networks through a grid of constraints, generating structure-preserving network embeddings. We use these embeddings to find clusters within the data that align both over time, as well as across languages.

The analysis of clusters across different languages can give us insights into how different language versions of RT make strategic use of tags to tailor information to their audiences and how this changes over time. We provide the first network analysis of the entire RT content based on tags assigned to the articles, thus, building a thematic network of the outlet. By providing a better understanding of the structure of RT and showing thematic differences across four language versions, our analysis highlights a lack of a coherent strategy in RT's targeting of audiences in different languages.

Firstly, we notice stark differences in the number of tags used and the way in which they cluster together across RT's language versions. 
Secondly, although there are clear overlaps between the networks of tags, large areas of clusters are unique for a given language version, pointing at the lack of an overarching idea for strategic communication with tags. 
Lastly, within language versions, we see significant shifts in the prioritization of certain tag clusters. 

Our findings are consistent with some of the previous studies on RT at an unprecedented scale. In particular, it has been shown that RT can express different political leanings \cite{elswah_anything_2020}, \cite{yablokov_conspiracy_2015} and its narratives are not necessarily coherent \cite{dajani_differentiated_2021}. The broadcaster rather adjusts the narratives based on the context and political situation. Thus, the observed variations in thematic clusters of tags may be evidence of that. 


\textbf{Limitations.}
Though we believe our method to be broadly applicable, it hinges upon several assumptions. Firstly, the order of alignments and constraints has to be hierarchical. In our case, the ties within one language are regarded as more important and act as the primary factor for network layout. Interlingual nodes can rearrange the network linearly but not change its inherent structure.
Secondly, there needs to be a sensible ratio between primary and secondary constraints for an adequate intermediate representation to exist. If the number of secondary constraints is too large, the intermediate representation must become very high-dimensional for good alignment performance. This diminishes the capability to retain structure from the primary constraints. Even if using a more intricate dimensionality reduction method than PCA, one must find a way to optimize for both sets of constraints at once or keep the intermediate representation low-dimensional. 
While interlingual alignment works reasonably well, it depends on the clustering step and hence does currently not improve the clustering performance. Leveraging BERT-Embeddings more thoroughly, future work might leverage the shared embedding space between languages to quantify semantically dense regions and areas of semantic divergence in time, as well as in language.
Due to the divergent use of tags by the Arabic, French, and Serbian language versions, our study can only show differences between the four languages. Moreover, the tags analysis does not represent topics in a strict sense and therefore cannot fully reflect RT's content orientation. Another limitation is that editorial boards of RT’s language versions likely have different approaches to the use of tags, which complicates the comparison. Future research may leverage the completeness of our collected dataset across all languages by performing topic modeling, sentiment analysis, or network analysis on full articles' titles and texts.

\printbibliography

@article{vavryk_mapping_2022,
	title = {Mapping Growth of the Russian Domestic Propaganda Apparatus on Telegram},
	volume = {2022},
	doi = {10.47459/cndcgs.2022.29},
	pages = {227--231},
	journaltitle = {Challenges to national defence in contemporary geopolitical situation},
	author = {Vavryk, Petro},
	date = {2022},
	note = {Publisher: Generolo Jono Žemaičio Lietuvos karo akademija},
}

@inbook{hutchings_dominant_2015,
	title = {Dominant Narratives in Russian Political and Media Discourse during the Ukraine Crisis},
	url = {https://www.e-ir.info/2015/04/28/dominant-narratives-in-russian-political-and-media-discourse-during-the-crisis/},
	booktitle = {Ukraine and Russia: People, Politics, Propaganda and Perspectives},
	publisher = {E-International Relations},
	author = {Hutchings, Stephen and Szostek, Joanna},
	bookauthor = {Pikulicka-Wilczewska, Agnieszka and Sakwa, Richard},
	date = {2015},
}

@article{crilley_emotions_2020,
	title = {Emotions and war on {YouTube}: affective investments in {RT}’s visual narratives of the conflict in Syria},
	volume = {33},
	issn = {0955-7571},
	url = {https://doi.org/10.1080/09557571.2020.1719038},
	doi = {10.1080/09557571.2020.1719038},
	pages = {713--733},
	number = {5},
	journaltitle = {Cambridge Review of International Affairs},
	shortjournal = {Cambridge Review of International Affairs},
	author = {Crilley, Rhys and Chatterje-Doody, Precious N.},
	date = {2020-09-02},
	note = {Publisher: Routledge},
}

@article{crilley_russia_2020,
	title = {From Russia with Lols: Humour, {RT}, and the Legitimation of Russian Foreign Policy},
	doi = {10.1080/13600826.2020.1839387},
	pages = {1--20},
	journaltitle = {Global Society},
	author = {Crilley, Rhys and Chatterje-Doody, Precious N.},
	date = {2020},
}

@report{helmus_russian_2018,
	location = {Santa Monica, {CA}},
	title = {Russian Social Media Influence: Understanding Russian Propaganda in Eastern Europe},
	institution = {{RAND} Corporation},
	author = {Helmus, Todd C. and Bodine-Baron, Elizabeth and Radin, Andrew and Magnuson, Madeline and Mendelonhn, Joshua and Marcellino, William and Bega, Andriy and Winkelman, Zev},
	date = {2018},
}

@inproceedings{vesselkov_russian_2020,
	location = {New York, {NY}, {USA}},
	title = {Russian Trolls Speaking Russian: Regional Twitter Operations and {MH}17},
	isbn = {978-1-4503-7989-2},
	url = {https://doi.org/10.1145/3394231.3397898},
	doi = {10.1145/3394231.3397898},
	series = {{WebSci} '20},
	pages = {86--95},
	booktitle = {Proceedings of the 12th {ACM} Conference on Web Science},
	publisher = {Association for Computing Machinery},
	author = {Vesselkov, Alexandr and Finley, Benjamin and Vankka, Jouko},
	date = {2020},
	note = {event-place: Southampton, United Kingdom},
}

@article{orttung_russia_2018,
	title = {Russia Today’s strategy and effectiveness on {YouTube}},
	volume = {35},
	pages = {77--92},
	number = {2},
	journaltitle = {Post-Soviet Affairs},
	author = {Orttung, Robert W. and Nelson, Elizabeth},
	date = {2018},
}

@article{bradshaw_disinformation_2019,
	title = {Disinformation optimised: gaming search engine algorithms to amplify junk news},
	volume = {8},
	doi = {10.14763/2019.4.1442},
	number = {4},
	journaltitle = {Internet Policy Review},
	author = {Bradshaw, Samantha},
	date = {2019},
}

@article{altay_misinformation_2023,
	title = {Misinformation on Misinformation: Conceptual and Methodological Challenges},
	volume = {9},
	issn = {2056-3051},
	url = {https://doi.org/10.1177/20563051221150412},
	doi = {10.1177/20563051221150412},
	pages = {20563051221150412},
	number = {1},
	journaltitle = {Social Media + Society},
	author = {Altay, Sacha and Berriche, Manon and Acerbi, Alberto},
	urldate = {2023-07-06},
	date = {2023-01-01},
	note = {Publisher: {SAGE} Publications Ltd},
}

@article{crilley_russia_2022,
	title = {‘Russia isn’t a country of Putins!’: How {RT} bridged the credibility gap in Russian public diplomacy during the 2018 {FIFA} World Cup},
	volume = {24},
	issn = {1369-1481},
	url = {https://doi.org/10.1177/13691481211013713},
	doi = {10.1177/13691481211013713},
	pages = {136--152},
	number = {1},
	journaltitle = {The British Journal of Politics and International Relations},
	author = {Crilley, Rhys and Gillespie, Marie and Kazakov, Vitaly and Willis, Alistair},
	urldate = {2023-07-06},
	date = {2022-02-01},
	note = {Publisher: {SAGE} Publications},
}

@article{dajani_differentiated_2021,
	title = {Differentiated visibilities: {RT} Arabic’s narration of Russia’s role in the Syrian war},
	volume = {14},
	issn = {1750-6352},
	url = {https://doi.org/10.1177/1750635219889075},
	doi = {10.1177/1750635219889075},
	shorttitle = {Differentiated visibilities},
	pages = {437--458},
	number = {4},
	journaltitle = {Media, War \& Conflict},
	author = {Dajani, Deena and Gillespie, Marie and Crilley, Rhys},
	urldate = {2023-07-06},
	date = {2021-12-01},
	langid = {english},
	note = {Publisher: {SAGE} Publications},
}

@article{muller_right-wing_2022,
	title = {Right-Wing, Populist, Controlled by Foreign Powers? Topic Diversification and Partisanship in the Content Structures of German-Language Alternative Media},
	volume = {10},
	issn = {2167-0811},
	url = {https://doi.org/10.1080/21670811.2022.2058972},
	doi = {10.1080/21670811.2022.2058972},
	pages = {1363--1386},
	number = {8},
	journaltitle = {Digital Journalism},
	shortjournal = {Digital Journalism},
	author = {Müller, Philipp and Freudenthaler, Rainer},
	date = {2022-09-14},
	note = {Publisher: Routledge},
}

@article{yuskiv_media_2021,
	title = {Media Reports as a Tool of Hybrid and Information Warfare (the Case of {RT} – Russia Today)},
	volume = {27},
	issn = {1224-032X},
	pages = {235--258},
	number = {1},
	journaltitle = {Codrul cosminului (Suceava, Romania)},
	author = {Yuskiv, Bohdan and Karpchuk, Nataliia and Khomych, Sergii},
	date = {2021},
	note = {Publisher: Stefan cel Mare University of Suceava},
}

@article{allcott_trends_2019,
	title = {Trends in the diffusion of misinformation on social media},
	volume = {6},
	rights = {The Author(s) 2019},
	issn = {2053-1680},
	pages = {1--8},
	number = {2},
	journaltitle = {Research \& Politics},
	author = {Allcott, Hunt and Gentzkow, Matthew and Yu, Chuan},
	date = {2019},
	note = {Place: London, England
Publisher: {SAGE} Publications},
}

@article{glazunova_soft_2022,
	title = {Soft power, sharp power? Exploring {RT}’s dual role in Russia’s diplomatic toolkit},
	volume = {0},
	issn = {1369-118X},
	url = {https://doi.org/10.1080/1369118X.2022.2155485},
	doi = {10.1080/1369118X.2022.2155485},
	shorttitle = {Soft power, sharp power?},
	pages = {1--26},
	number = {0},
	journaltitle = {Information, Communication \& Society},
	author = {Glazunova, Sofya and Bruns, Axel and Hurcombe, Edward and Montaña-Niño, Silvia Ximena and Coulibaly, Souleymane and Obeid, Abdul Karim},
	urldate = {2023-07-05},
	date = {2022-12-20},
	note = {Publisher: Routledge
\_eprint: https://doi.org/10.1080/1369118X.2022.2155485},
}

@collection{woolley_computational_2019,
	location = {New York, {NY}},
	title = {Computational propaganda : political parties, politicians, and political manipulation on social media},
	isbn = {0-19-093143-4},
	series = {Oxford studies in digital politics},
	publisher = {Oxford University Press},
	editor = {Woolley, S. and Howard, P.N},
	date = {2019},
}

@article{howard_algorithms_2018,
	title = {Algorithms, Bots, and Political Communication in the {US} 2016 Election: The challenge of Automated Political Communication for Election Law and Administration},
	volume = {15},
	doi = {https://doi.org/10.1080/19331681.2018.1448735},
	pages = {81--91},
	number = {2},
	journaltitle = {Journal of Information Technology \& Politics},
	author = {Howard, P.N and Woolley, S. and Calo, R.},
	date = {2018},
}

@article{muhammed_disaster_2022,
	title = {The disaster of misinformation: a review of research in social media. International journal of data science and analytics},
	volume = {13},
	doi = {https://doi.org/10.1007/s41060-022-00311-6},
	pages = {271--285},
	number = {4},
	journaltitle = {International journal of data science and analytics},
	author = {Muhammed, T.S. and Mathew, S.K.},
	date = {2022},
}

@article{blumler_fourth_2016,
	title = {The Fourth Age of Political Communication},
	volume = {6},
	pages = {19--30},
	number = {1},
	journaltitle = {Politiques de communication},
	author = {Blumler, Jay G.},
	date = {2016},
}

@article{kuznetsova_kontrpropaganda_2021,
	title = {Kontrpropaganda today: The roots of {RT}’s defensive practices and countering ethic},
	issn = {1464-8849, 1741-3001},
	url = {http://journals.sagepub.com/doi/10.1177/14648849211033442},
	doi = {10.1177/14648849211033442},
	pages = {146488492110334},
	journaltitle = {Journalism},
	shortjournal = {Journalism},
	author = {Kuznetsova, Elizaveta},
	urldate = {2023-03-03},
	date = {2021-08-13},
	langid = {english},
}

@article{elswah_anything_2020,
	title = {“Anything that Causes Chaos”: The Organizational Behavior of Russia Today ({RT})},
	volume = {70},
	issn = {0021-9916, 1460-2466},
	url = {https://academic.oup.com/joc/article/70/5/623/5912109},
	doi = {10.1093/joc/jqaa027},
	shorttitle = {“Anything that Causes Chaos”},
	pages = {623--645},
	number = {5},
	journaltitle = {Journal of Communication},
	author = {Elswah, Mona and Howard, Philip N},
	urldate = {2023-01-27},
	date = {2020-10-01},
	langid = {english},
}

@article{hoyle_portrait_2021,
	title = {Portrait of liberal chaos: {RT}’s antagonistic strategic narration about the Netherlands},
	issn = {1750-6352, 1750-6360},
	url = {http://journals.sagepub.com/doi/10.1177/17506352211064705},
	doi = {10.1177/17506352211064705},
	shorttitle = {Portrait of liberal chaos},
	pages = {175063522110647},
	journaltitle = {Media, War \& Conflict},
	shortjournal = {Media, War \& Conflict},
	author = {Hoyle, Aiden and van den Berg, Helma and Doosje, Bertjan and Kitzen, Martijn},
	urldate = {2023-01-27},
	date = {2021-12-29},
	langid = {english},
}

@article{yablokov_conspiracy_2015,
	title = {Conspiracy Theories as a Russian Public Diplomacy Tool: The Case of Russia Today ({RT})},
	volume = {35},
	issn = {0263-3957},
	url = {https://doi.org/10.1111/1467-9256.12097},
	doi = {10.1111/1467-9256.12097},
	shorttitle = {Conspiracy Theories as a Russian Public Diplomacy Tool},
	pages = {301--315},
	number = {3},
	journaltitle = {Politics},
	author = {Yablokov, Ilya},
	urldate = {2023-07-04},
	date = {2015-11-01},
	langid = {english},
	note = {Publisher: {SAGE} Publications Ltd},
}

@article{bazziCommunityDetectionTemporal2016,
  title = {Community {{Detection}} in {{Temporal Multilayer Networks}}, with an {{Application}} to {{Correlation Networks}}},
  author = {Bazzi, Marya and Porter, Mason A. and Williams, Stacy and McDonald, Mark and Fenn, Daniel J. and Howison, Sam D.},
  year = {2016},
  month = jan,
  journal = {Multiscale Model. Simul.},
  volume = {14},
    pages = {1--41},
        langid = {english}
}

@article{bonifaziInvestigatingCOVID19Vaccine2022,
  title = {Investigating the {{COVID-19}} Vaccine Discussions on {{Twitter}} through a Multilayer Network-Based Approach},
  author = {Bonifazi, Gianluca and Breve, Bernardo and Cirillo, Stefano and Corradini, Enrico and Virgili, Luca},
  year = {2022},
  month = nov,
  journal = {Information Processing \& Management},
  volume = {59},
    pages = {103095},
        langid = {english}
}

@article{dedomenicoMathematicalFormulationMultilayer2013,
  title = {Mathematical {{Formulation}} of {{Multilayer Networks}}},
  author = {De Domenico, Manlio and {Sol{\'e}-Ribalta}, Albert and Cozzo, Emanuele and Kivel{\"a}, Mikko and Moreno, Yamir and Porter, Mason A. and G{\'o}mez, Sergio and Arenas, Alex},
  year = {2013},
  month = dec,
  journal = {Phys. Rev. X},
  volume = {3},
    pages = {041022},
        langid = {english}
}

@article{ghawiCommunityMatchingBased2022,
  title = {A Community Matching Based Approach to Measuring Layer Similarity in Multilayer Networks},
  author = {Ghawi, Raji and Pfeffer, J{\"u}rgen},
  year = {2022},
  month = jan,
  journal = {Social Networks},
  volume = {68},
  pages = {1--14},
  langid = {english}
}

@article{mcinnesHdbscanHierarchicalDensity2017,
  title = {Hdbscan: {{Hierarchical}} Density Based Clustering},
  shorttitle = {Hdbscan},
  author = {McInnes, Leland and Healy, John and Astels, Steve},
  year = {2017},
  month = mar,
  journal = {The Journal of Open Source Software},
  volume = {2}
}

@article{oroDetectingTopicAuthoritative2018,
  title = {Detecting {{Topic Authoritative Social Media Users}}: {{A Multilayer Network Approach}}},
  shorttitle = {Detecting {{Topic Authoritative Social Media Users}}},
  author = {Oro, Ermelinda and Pizzuti, Clara and Procopio, Nicola and Ruffolo, Massimo},
  year = {2018},
  month = may,
  journal = {IEEE Trans. Multimedia},
  volume = {20},
    pages = {1195--1208},
      }

@article{pyoNetworkPropagandaManipulation2019,
  title = {Network {{Propaganda}}: {{Manipulation}}, {{Disinformation}}, and {{Radicalization}} in {{American Politics}}},
  shorttitle = {Network {{Propaganda}}},
  author = {Pyo, Yeahin Jane},
  year = {2019},
  month = jan,
  journal = {International journal of communication (Online)},
  pages = {462--465},
  langid = {english}
}

@article{salamaTemporalNetworksReview2022,
  title = {Temporal Networks: A Review and Opportunities for Infrastructure Simulation},
  shorttitle = {Temporal Networks},
  author = {Salama, Mohamed and Ezzeldin, Mohamed and {El-Dakhakhni}, Wael and Tait, Michael},
  year = {2022},
  month = jan,
  journal = {Sustainable and Resilient Infrastructure},
  volume = {7},
    pages = {40--55},
        langid = {english}
}

@article{borgattiNetworkAnalysisSocial2009,
  title = {Network {{Analysis}} in the {{Social Sciences}}},
  author = {Borgatti, Stephen P. and Mehra, Ajay and Brass, Daniel J. and Labianca, Giuseppe},
  year = {2009},
  month = feb,
  journal = {Science},
  volume = {323},
    pages = {892--895}
}

@article{daduApplicationAlignedUMAPLongitudinal2023,
  title = {Application of {{Aligned-UMAP}} to Longitudinal Biomedical Studies},
  author = {Dadu, Anant and Satone, Vipul K. and Kaur, Rachneet and Koretsky, Mathew J. and Iwaki, Hirotaka and Qi, Yue A. and Ramos, Daniel M. and Avants, Brian and Hesterman, Jacob and Gunn, Roger and Cookson, Mark R. and Ward, Michael E. and Singleton, Andrew B. and Campbell, Roy H. and Nalls, Mike A. and Faghri, Faraz},
  year = {2023},
  month = jun,
  journal = {Patterns},
  volume = {4},
    pages = {100741},
  langid = {english}
}

@article{delfresnogarciaIdentifyingNewInfluences2016,
  title = {Identifying the New {{Influences}} in the {{Internet Era}}: {{Social Media}} and {{Social Network Analysis}}: {{Identificando}} a Los Nuevos Influyentes En Tiempos de {{Internet}}: Medios Sociales y An\'alisis de Redes Sociales.},
  shorttitle = {Identifying the New {{Influences}} in the {{Internet Era}}},
  author = {{del Fresno Garc{\'i}a}, Miguel and Daly, Alan J. and {Segado S{\'a}nchez-Cabezudo}, Sagrario},
  year = {2016},
  month = jan,
  journal = {Revista Espa\~nola de Investigaciones Sociologicas},
    pages = {23--40}
}

@misc{devlinBERTPretrainingDeep2019,
  title = {{{BERT}}: {{Pre-training}} of {{Deep Bidirectional Transformers}} for {{Language Understanding}}},
  shorttitle = {{{BERT}}},
  author = {Devlin, Jacob and Chang, Ming-Wei and Lee, Kenton and Toutanova, Kristina},
  year = {2019},
  month = may,
  archiveprefix = {arxiv}
}

@misc{ghoniemStateArtMultilayer2019,
  title = {The {{State}} of the {{Art}} in {{Multilayer Network Visualization}}},
  author = {Ghoniem, Mohammad and Mcgee, Fintan and Melan{\c c}on, Guy and Otjacques, Benoit and Pinaud, Bruno},
  year = {2019},
  month = feb,
  archiveprefix = {arxiv}
}

@misc{mcinnesUMAPUniformManifold2020,
  title = {{{UMAP}}: {{Uniform Manifold Approximation}} and {{Projection}} for {{Dimension Reduction}}},
  shorttitle = {{{UMAP}}},
  author = {McInnes, Leland and Healy, John and Melville, James},
  year = {2020},
  month = sep,
  archiveprefix = {arxiv}
}

@misc{rahimiANTMAlignedNeural2023,
  title = {{{ANTM}}: {{An Aligned Neural Topic Model}} for {{Exploring Evolving Topics}}},
  shorttitle = {{{ANTM}}},
  author = {Rahimi, Hamed and Naacke, Hubert and Constantin, Camelia and Amann, Bernd},
  year = {2023},
  month = jun,
  archiveprefix = {arxiv}
}
\end{document}